\documentclass[a4paper,12pt]{article}
\usepackage{amsmath,amssymb,latexsym}
\usepackage[dvips]{graphicx}

\newcommand{\xhat}{\hat{x}}

\newcommand{\stilde}{\tilde{\sigma}}

\newcommand{\xvec}{\vec{x}}
\newcommand{\xpvec}{\vec{x'}}
\newcommand{\xppvec}{\vec{x''}}

\newcommand{\yvec}{\vec{y}}
\newcommand{\ypvec}{\vec{y'}}

\newcommand{\zvec}{\vec{z}}
\newcommand{\zpvec}{\vec{z'}}

\newcommand{\avec}{\vec{a}}
\newcommand{\bvec}{\vec{b}}
\newcommand{\cvec}{\vec{c}}

\begin{document}
\renewcommand{\thefootnote}{\fnsymbol{footnote}}
\begin{titlepage}
\hfill
{\hfill \begin{flushright} 
YITP-09-10
\end{flushright}
  }

\vspace*{10mm}

\begin{center}
{\LARGE {\LARGE  
Massive particles coupled with 2+1 dimensional gravity and noncommutative field theory}}
\vspace*{18mm}

 {\large Yuya Sasai\footnote{ e-mail: sasai@yukawa.kyoto-u.ac.jp}
 and Naoki Sasakura\footnote{e-mail: sasakura@yukawa.kyoto-u.ac.jp}}

\vspace*{13mm}
{\large {\it 
Yukawa Institute for Theoretical Physics, Kyoto University, \\ 
 Kyoto 606-8502, Japan 
}} \\

\end{center}

\vspace*{15mm}

\begin{abstract}
Recently, it has been shown that the effective field theory of the Ponzano-Regge model with which spinless massive particles are coupled   is given by three dimensional Euclidean noncommutative scalar field theory in the Lie algebraic noncommutative space $[\xhat^i, \xhat^j]=2i\kappa \epsilon^{ijk}\xhat_k ~(i,j,k=1,2,3)$ with $\kappa=4\pi G$, where $G$ is a gravitational constant.
We examine whether there exists the relation between spinless massive particles coupled with $2+1$ dimensional Einstein gravity and  the Lorentzian version of the noncommutative  field theory.
Then, we point out that the momentum space of the spinless  massive particles in 2+1 dimensional Einstein gravity is generally different from that of the noncommutative field theory, which is given by $SL(2,R)/Z_2$ group space.
\end{abstract}

\end{titlepage}

\newpage
\renewcommand{\thefootnote}{\arabic{footnote}}
\setcounter{footnote}{0}

\section{Introduction}
Quantum gravity is one of the goals of high energy physics. Now, we have some promising approaches to understand quantum gravity, such as string theory and loop quantum gravity. But to study simple models of gravity is also a hopeful direction to reveal the realistic quantum gravity. For example, three dimensional gravity is simpler than our universe because there is no gravitational degree of freedom in three dimensions. But if a massive particle is coupled with three dimensional gravity, the geometry is deformed conically \cite{Deser:1983tn}. Thus, it might be interesting if we consider the quantization of such massive particles because the background geometry itself seems to be fluctuated.

So far, many works about the quantization have been done. In \cite{'tHooft:1988yr,Deser:1988qn,de Sousa Gerbert:1988bj,Matschull:1997du,Louko:1999hk,Matschull:2001ec,Louko:2001ed}, quantum mechanics of the massive particles coupled with three dimensional Einstein gravity has been considered. In \cite{Bais:1998yn,Bais:2002ye}, the authors have quantized the moduli space of three dimensional Chern-Simons gauge theory coupled with particles and derived a general expression for the scattering cross section of the gravitating particles. In \cite{Meusburger:2003ta,Meusburger:2003hc,Meusburger:2005mg}, more general moduli space of three dimensional Chern-Simons gauge theory have been considered. These works have contributed to the first quantization of the massive particles in three dimensions.  

The field theoretical description of the massive particles in three dimensions has appeared in \cite{Freidel:2004vi,Freidel:2005bb}. The authors have considered the Ponzano-Regge model \cite{Ponzano:1968} with which spinless massive particles are coupled and have shown that the effective field theory of the Ponzano-Regge model is given by three dimensional noncommutative scalar field theory in the Lie algebraic noncommutative space $[\xhat^i, \xhat^j]=2i\kappa \epsilon^{ijk}\xhat_k ~(i,j,k=1,2,3)$ with $\kappa=4\pi G$, where $G$ is a gravitational constant. This result means that the three dimensional noncommutative field theory in the Lie algebraic noncommutative space might give the quantum dynamics of spinless massive particles coupled with the three dimensional gravity. If this conjecture is correct, it is natural to expect that the Lorentzian extension of the noncommutative field theory gives the quantum dynamics of spinless massive particles in 2+1 dimensional Einstein gravity.

In this paper, we examine whether there exists  the relation between spinless massive particles coupled with $2+1$ dimensional Einstein gravity and the Lorentzian version of the noncommutative field theory. It is known that the 2+1 dimensional noncommutative  field theory in the Lie algebraic noncommutative spacetime $[\xhat^i, \xhat^j]=2i\kappa \epsilon^{ijk}\xhat_k ~(i,j,k=0,1,2)$  possesses an $SL(2,R)/Z_2$ group momentum space \cite{Sasakura:2000vc,Imai:2000kq,Sasai:2007mc}. Since it has been reported in \cite{Matschull:1997du,Bais:2002ye,Meusburger:2003ta} that  spinless massive particles coupled with the 2+1 dimensional gravity possess an $SL(2,R)/Z_2$ group momentum space, the noncommutative field theory seems to give the quantum dynamics of such spinless massive particles.
However, we point out that full momentum space of the spinless massive particles is generally different from $SL(2,R)/Z_2$ group space because there are arbitrary negative energy particle solutions in the 2+1 dimensional gravity.
In fact, we show that the statement  ``spinless massive particles in 2+1 dimensional Einstein gravity possess an $SL(2,R)/Z_2$ group momentum space" is correct only when  masses of the particles are positive and the total energy is not over $1/4G$.

This paper is organized as follows. In section \ref{subsec:static}, we review a static spinless massive particle solution in 2+1 dimensional Einstein gravity. In section \ref{subsec:geo}, we review the geometric approach \cite{Deser:1983tn} which gives the way to construct the spinless massive moving particles in 2+1 dimensions. In section \ref{subsec:gmomenta}, we point out that the momentum space of the spinless massive particles can be written by the $SL(2,R)/Z_2$ group space only when  masses of the particles are positive and the total energy is not over $1/4G$.
 The final section is devoted to a summary and comments.

\section{Particles coupled with 2+1 dimensional gravity and the momentum space} \label{sec:3dqg}
\subsection{A static particle in 2+1 dimensional Einstein gravity} \label{subsec:static}
Let us remind  a static spinless massive particle solution in 2+1 dimensional Einstein gravity. The metric is given by
\begin{equation}
ds^2=-dt^2+dr^2+r^2d\phi^2,
\end{equation}
where the range of $\phi$ is $0\leq \phi \leq 2\pi (1-4Gm)$ and $m$ is the mass of the particle. Since this metric is locally flat, we can obtain the classical solution if we start with flat Minkowski space, cut out a wedge with opening angle $2\beta\equiv 8\pi Gm$, and identify the opposite edges.
Since  $\beta$ should be less than $\pi$, $m$ has an upper bound,
\begin{equation}
m \leq \frac{1}{4G}. \label{eq:massbound}
\end{equation}
However, arbitrary negative values of $m$ are permitted because there is no lower bound of $\beta$. Thus, a particle coupled with 2+1 dimensional Einstein gravity can possess arbitrary negative mass.

\subsection{Geometric approach} \label{subsec:geo}
To consider moving massive particles coupled with 2+1 dimensional Einstein gravity, it is convenient to introduce the geometrical approach \cite{Deser:1983tn}.

\begin{figure}
\begin{center}
\includegraphics[scale=0.8]{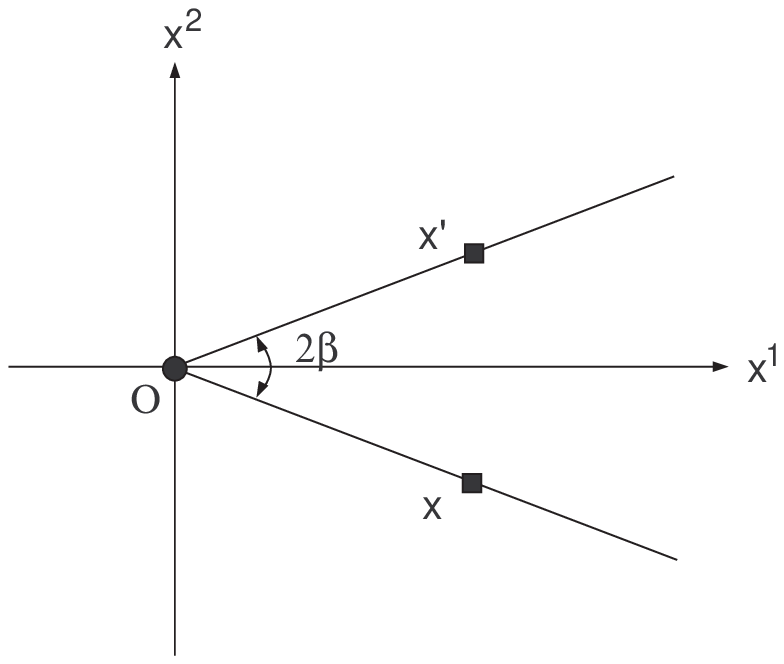}
\end{center}
\caption{Two dimensional space when there is a static spinless massive particle at the origin. The mark $\blacksquare$ means the identification of the points $x^i$ and $x^{'i}$.}
\label{fig:static1}
\end{figure}

Let us reconsider the static spinless massive one-particle solution. Such a solution is determined by the matching condition which is expressed by identifying points $x^i$ and $x^{'i}$ along the edges which are related by the rotation matrix $\Omega$ as in Figure \ref{fig:static1},

\begin{align}
\xpvec&=\Omega(\beta) \xvec, \notag \\
\Omega(\beta)&\equiv
\begin{pmatrix}
1 & 0 & 0 \\
0 & ~\cos 2\beta & ~\sin 2\beta \\
0 & ~-\sin 2\beta & ~\cos 2\beta 
\end{pmatrix},
\end{align}
where $\xvec$ denotes the three vector. In this setting, however, $\beta$ should be restricted to a positive value because of the periodic property of $\Omega(\beta)$. Of course, $\beta$ can not be taken over $\pi$.

Next, let us consider the two-particle static solution. The matching conditions for two particles at the origin and at $\avec=(0,a,0)$ consist in the following identifications as in Figure \ref{fig:static2}:
\begin{figure}
\begin{center}
\includegraphics[scale=0.7]{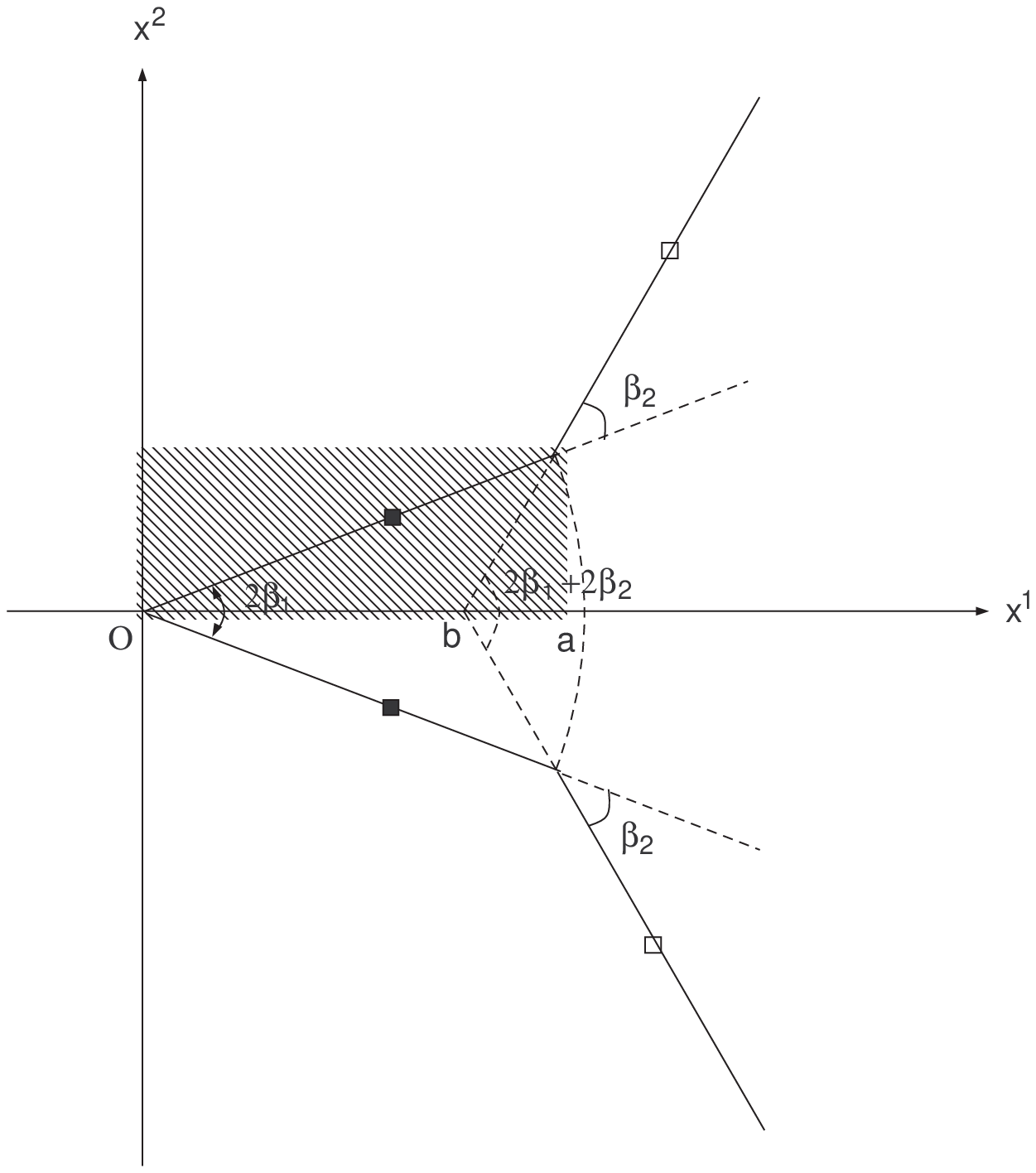}
\end{center}
\caption{Two dimensional space when there are two static spinless massive particles at the origin and at $\avec=(0,a,0)$. The mark $\blacksquare$ and $\square$ mean the identifications.}
\label{fig:static2}
\end{figure}
\begin{align}
\xvec \sim \xpvec &=\Omega_1\xvec, \\
\xvec \sim \xppvec &=\Omega_1\{\avec+\Omega_2(\xvec-\avec)\} \notag \\
&=\bvec+\Omega_1\Omega_2(\xvec-\bvec), \label{eq:2pid}
\end{align}
where 
\begin{equation}
\bvec=\frac{\sin\beta_2}{\sin(\beta_1+\beta_2)}\Omega_1^{1/2}\avec, ~~~\Omega_{1,2}\equiv \Omega(\beta_{1,2}).
\end{equation}
We see that the identification (\ref{eq:2pid}) coincides with a one-particle solution with mass $m'=m_1+m_2$ located at $x^i=b^i$. Thus, the total mass should be bounded by $m_1+m_2<1/4G$ so that spacetime with Minkowskian metric exists.

Finally, we consider two moving particles. Since we can not discuss a total momentum of an isolated system, we should study two moving particles in the center-of-mass frame.

\begin{figure}
\begin{center}
\includegraphics[scale=0.8]{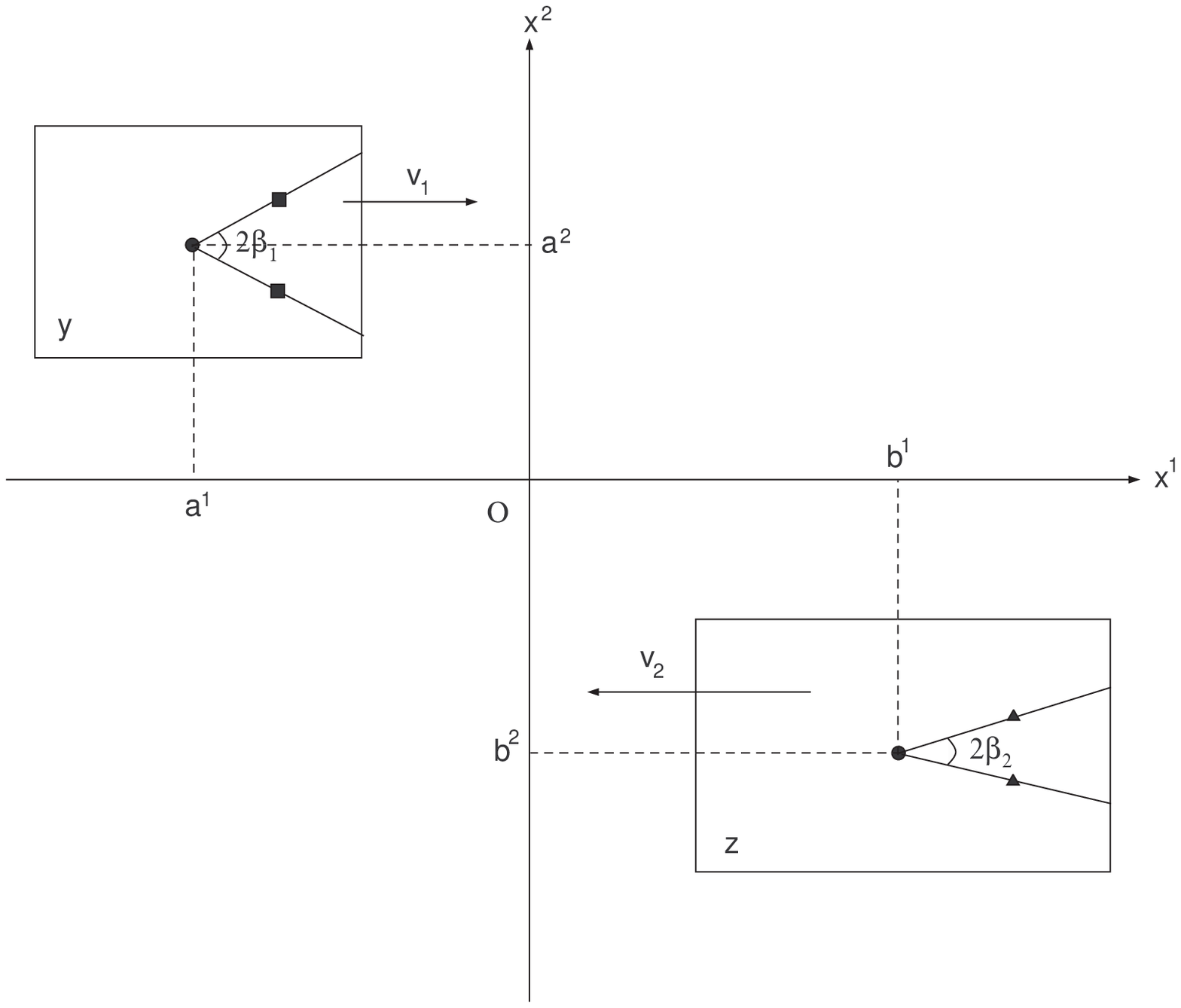}
\end{center}
\caption{The embedding of the two patches.}
\label{fig:movingparticles}
\end{figure}

Let us prepare two local patches whose coordinates are labeled as $y^i$ and $z^i$ respectively. If a static particle is in each patch, the matching conditions are given by
\begin{align}
\yvec \sim \ypvec &=\Omega_1\yvec, \label{eq:yidentify} \\
\zvec \sim \zpvec &=\Omega_2\zvec. \label{eq:zidentify}
\end{align}
Next, we consider the embedding of these two patches into a global frame whose coordinates are labeled by $x^i$ as in Figure \ref{fig:movingparticles}. 
If these particles are moving and located at $\avec=(0,a^1,a^2)$ and $\bvec=(0,b^1,b^2)$ in the $x$ frame, the relations between these coordinates are 
\begin{align}
\xvec&=\Lambda_1\yvec+\avec \label{eq:xyrel} \\
&=\Lambda_2\zvec+\bvec, \label{eq:xzrel}
\end{align}
where $\Lambda_{1,2}$ are the Lorentz transformations with the boost parameters $\gamma_{1,2}$. If a particle is moving in $x^1$ direction, the Lorentz transformation $\Lambda$ is written by
\begin{align}
\Lambda&=
\begin{pmatrix}
\cosh \gamma & ~\sinh \gamma & ~0 \\
\sinh \gamma & ~\cosh \gamma & ~0 \\
0 & 0 & ~1 
\end{pmatrix}, \notag \\
&\tanh \gamma =v,
\end{align}
where $v$ is a velocity of the particle.
From (\ref{eq:yidentify})-(\ref{eq:xzrel}), the matching conditions for two particles moving with the velocity $v_{1,2}$ respectively are the following,
\begin{align}
\xvec \sim \xpvec &=\avec+\Lambda_1\Omega_1\Lambda_1^{-1}(\xvec-\avec), \label{eq:movingid} \\
\xvec \sim \xppvec &=\avec+\Lambda_1\Omega_1\Lambda_1^{-1}(\bvec-\avec+\Lambda_2\Omega_2\Lambda_2^{-1}(\xvec-\bvec)). \label{eq:centermass1}
\end{align}
Note that the matching condition for $x^i$, which is far from two particles, is given by (\ref{eq:centermass1}).

If we are in the center mass frame, the matching condition should be written as \begin{equation}
\xvec\sim \xppvec=\Omega_3\xvec+\cvec, \label{eq:centermass2}
\end{equation}
where $\Omega_3$ is purely spacelike and $\cvec$ is a three vector. Comparing (\ref{eq:centermass1}) and (\ref{eq:centermass2}), we obtain
\begin{equation}
\Omega_3=\Lambda_1\Omega_1\Lambda_1^{-1}\Lambda_2\Omega_2\Lambda_2^{-1}. \label{eq:comtotalE}
\end{equation}
From (\ref{eq:comtotalE}), we find the total energy of the center-of-mass frame and the conditions for the center-of-mass frame.

\subsection{Momentum space of massive particles in 2+1 dimensions and $SL(2,R)/Z_2$ group space} \label{subsec:gmomenta}
In this section, we show that momentum space of spinless massive particles coupled with 2+1 dimensional Einstein gravity is given by an $SL(2,R)/Z_2$ group space only when their masses are positive and the total energy is not over $1/4G$.

We can represent (\ref{eq:comtotalE}) by $SL(2,R)/Z_2$ group elements because $SO(2,1)$ is isomorphic to $SL(2,R)/Z_2$. If $\Lambda$ represents the boost in $x^1$ direction, $\Lambda\Omega\Lambda^{-1}$ is written with $SO(2,1)$ Lie algebras $T_i$ as
\begin{align}
\Lambda\Omega\Lambda^{-1}&=e^{i\gamma T_2}e^{i2\beta T_0}e^{-i\gamma T_2} \notag \\
&=e^{i2\beta (\cosh \gamma T_0+\sinh \gamma T_1)}, \label{eq:lol-1}
\end{align}
where
\begin{equation}
T_0=
\begin{pmatrix}
0 & ~~0 & ~0 \\
0 & ~~0 & ~-i \\
0 & ~~i & ~0
\end{pmatrix}
, ~T_1=
\begin{pmatrix}
0 & ~0 & ~-i \\
0 & ~0 & ~0 \\
-i & ~0 & ~0
\end{pmatrix}
, ~T_2=
\begin{pmatrix}
0 & ~-i & ~0 \\
-i & ~0 & ~0 \\
0 & ~0 & ~0
\end{pmatrix},
\end{equation}
which satisfy the following commutation relation:\footnote{The signatures of the metric and the totally antisymmetric tensor are the following:
\begin{align}
\eta^{ij}&=(-1,1,1), \notag \\
\epsilon^{012}&=1. \notag
\end{align}} 
\begin{equation}
[T^i,T^j]=i\epsilon^{ijk}T_k.
\end{equation}

Replacing $T^i$ with $SL(2,R)$ Lie algebras $\stilde^i/2$,\footnote{The $\tilde{\sigma}^i$s are defined by \[\tilde{\sigma}^0=
\sigma^2, ~\tilde{\sigma}^1=i\sigma^3,~ \tilde{\sigma}^2=i\sigma^1,\] with Pauli 
matrices
\[
\sigma^1=
\begin{pmatrix}
0 & ~1 \\
1 & ~0 \\
\end{pmatrix},~
\sigma^2=
\begin{pmatrix}
0 & ~-i \\
i & ~0 \\ 
\end{pmatrix},~
\sigma^3=
\begin{pmatrix}
1 & 0 \\
0 & ~-1 \\
\end{pmatrix}.
\]
 These matrices satisfy
\[
 \stilde^i\stilde^j=-\eta^{ij}+i\epsilon^{ijk}\stilde_k.
\]
} (\ref{eq:lol-1}) can be represented as
\begin{equation}
g=e^{i\kappa k\cdot \stilde}, \label{eq:gexp}
\end{equation}
with the identification $g \sim -g$.
If we set 
\begin{equation}
\kappa =4\pi G, \label{eq:relationofkappa}
\end{equation}
we find
\begin{align}
k_0=m\cosh \gamma, ~~~k_1=m\sinh \gamma, ~~~k_2=0.
\end{align}
This seems to allow us to interpret $k_i$ as relativistic momenta of a massive particle in 2+1 dimensional spacetime. 
Then, we can obtain the total energy of two massive particles in the center-of-mass frame from the product of $SL(2,R)/Z_2$ group elements in the same way as we have done in (\ref{eq:comtotalE}). 
But we must remember that this discussion is only valid when masses of the particles are positive and the total energy is not over $1/4G$. Thus, we have found that the momentum space of spinless massive particles coupled with 2+1 dimensional Einstein gravity is given by the $SL(2,R)/Z_2$ group space only when their masses are positive and the total energy is not over $1/4G$.

Since the $SL(2,R)/Z_2$ group momentum space does not possess an arbitrary negative energy, this momentum space does not correspond to the negative energy particle solutions as we saw in section \ref{subsec:static}.  Thus, full momentum space of the spinless massive particles in 2+1 dimensional Einstein gravity is much more complicated than  $SL(2,R)/Z_2$ group space.

\section{Summary and comments}
We have shown by using the geometric approach  that spinless massive particles in 2+1 dimensional Einstein gravity possess the $SL(2,R)/Z_2$ group momentum space only when the masses of the particles are positive and the total energy is not over $1/4G$. This result tells us that the $SL(2,R)/Z_2$ group space does not give the full momentum space of the particles in the 2+1 dimensional gravity. In fact, the $SL(2,R)/Z_2$ group momentum space does not contain the arbitrary negative energy which the particles  should possess.

On the other hand, it is known that 2+1 dimensional noncommutative field theory in the Lie algebraic noncommutative spacetime $[\xhat^i, \xhat^j]=2i\kappa \epsilon^{ijk}\xhat_k ~(i,j,k=0,1,2)$, which is  a Lorentzian extension of the effective field theory of the Ponzano-Regge model with scalar massive particles, possesses the $SL(2,R)/Z_2$ group momentum space. However, since  this theory does not contain the arbitrary negative energy, we conclude that the noncommutative field theory in the Lie algebraic noncommutative spacetime is not related to the spinless massive particles coupled with 2+1 dimensional gravity. 

Two comments are in order. Firstly, we have recently shown that the noncommutative field theory in the Lie algebraic noncommutative spacetime  is not probably a unitary theory \cite{Sasai:2009jm}. If there is a true quantum gravity theory, it should be consistent with the principle of the quantum mechanics. From this result, it is  unlikely that the noncommutative field theory is the effective field theory of the  spinless massive particles coupled with the 2+1 dimensional gravity. 

Secondly, what is the full momentum space of the spinless massive particles in the  2+1 dimensional gravity? Since the periodic property of $SL(2,R)/Z_2$ group forbids the arbitrary negative energy, the universal covering of the $SL(2,R)/Z_2$  seems to be considered as a candidate of the full momentum space of the particles. However, in this case, we do not know how to understand the upper bound of the total energy. This problem should remain as a future work.

\section*{Acknowledgments}
Y.S. was supported in part by JSPS Research Fellowships for Young Scientists.
N.S. was supported in part by the Grant-in-Aid for Scientific Research No.~18340061 from the Ministry of Education, Science, Sports and Culture of Japan.

\end{document}